\documentstyle[12pt]{article}
\advance\textheight by \topskip
\begin{document}
\begin{flushright}
SU-ITP-96-58\\
hep-th/9612076\\
December 6, 1966\\
\end{flushright}
\vspace{1cm}
\begin{center}
\baselineskip=16pt

{\Large\bf   CALABI-YAU BLACK HOLES }  \\

\vskip 2cm

{\bf  Marina Shmakova\footnote { E-mail:
shmakova@slac.stanford.edu} }\\
 \vskip 0.8cm
University of Tennessee, Knoxville, TN 37996\\
Stanford Linear Accelerator Center, Stanford University,
Stanford, CA 94309\\
\vskip .6cm

\vskip 1 cm

\end{center}
\vskip 1 cm
\centerline{\bf ABSTRACT}
\begin{quotation}
We  have found  the entropy of  N=2  extreme  black holes
associated with  general Calabi-Yau moduli space and prepotential
$F= d_{ABC}{X^A X^B X^C\over X^0} $. We show that for arbitrary
$d_{ABC}$
and  black hole charges  $p$ and $q$ the entropy-area formula depends on
combinations
of these charges and parameters  $d_{ABC}$. These combinations  are the
solutions of the simple system of algebraic equations.
We gave a few examples of particular Calabi-Yau moduli space for
which this
system has an explicit  solution.
For special case when one of black hole charges is equal to zero
($p^0=0$)
the solution always exists.
 \end{quotation}
\newpage

\section{Introduction}

	Recently supersymmetric black holes become a
``natural'' theoretical laboratory for string theory.
 Calculation of black hole entropy by counting microscopic configurations
\cite{SV,KLMS,BM}  and comparison of these
results with
 one obtained by classical macroscopic calculations
\cite{FK,FKS}  is a
very powerful tool in this  ``laboratory''.  In this paper we will
focus on
macroscopic calculations of the black hole entropy as the area of
the black
hole horizon.  We will give a solution for  N=2 extreme black holes
associated
with the general Calabi-Yau moduli space.

One of the most important  properties
of  N=2
supersymmetric black holes
in the general version of ungauged N=2
supergravity interacting  with arbitrary number $n_v$ of   vector
multiplets \cite{Cer}
 has been established recently \cite{FKS}. It was shown
that the area of
the extreme black hole horizon and
moduli of vector
multiplets near the horizon are functions of charges only.
 \begin{equation}
A =A( z^i(p,q, d_{ijk}),p,q, d_{ijk}),
\label{extr}\end{equation}
where  $ A $ is an area of the black hole horizon, $ z^i(p,q,
d_{ijk}) $ are
moduli fields, $p $ and $ q $ are electrical and magnetic charges
of black hole
and $ d_{ijk} $
is a characteristic of Calabi-Yau (classical intersection numbers).
Relation
between charges and moduli for supersymmetric black
holes near the horizon  has the  following form \cite{FK}:
\begin{equation}
\left (\matrix{
p^\Lambda\cr
q_\Lambda\cr
}\right )={ \rm Re} \left ( \matrix{
2i \bar Z L^\Lambda\cr
2i \bar Z M_\Lambda\cr
}\right ) \ ,
\label{stab}\end{equation}
where $Z$ is the central charge  and ($L^\Lambda , M_\Lambda $) are
covariantly
holomorphic sections.

Classical Calabi-Yau moduli space is
described by the
prepotential of the form:
\begin{equation}
F= d_{ABC}{X^A X^B X^C\over X^0}+\quad corrections.
\label{ppd}\end{equation}
Here $d_{ABC}$ are the topological intersection numbers
$d_{ABC}\equiv \int
J_A\wedge J_B\wedge J_C $ and $ J_A $ K\"ahler cone generators, $ J_A \in
H^{1,1}(Y,{\bf Z})$. Special coordinates $z^A$ of   K\"ahler  ``moduli
space'' are
connected with $X^\Lambda $ by:
\begin{equation}
z^A={X^A \over X^0}; \qquad  X^0=1; \quad A=1,2... ,\quad \Lambda
=0,1... .
\end{equation}
 According to
\cite{HLY,LSTY,BCWKLM} this form of the  Calabi-Yau
prepotential
can be extended by addition  of an extra topological term
determined by second
Chern class $c_2$  to the  prepotential:
\begin{equation}
F_1(z)=\sum_A^h {c_2\cdot J_A\over 24}z^A.
\label{ppdc}\end{equation}
Symplectic invariant form of the  K\"ahler potential is $K=-\ln i( \bar
X^\Lambda
F_\Lambda  -  X^\Lambda \bar F_\Lambda  ) $, where
\begin{equation}
L^\Lambda=e^{K/2}X^\Lambda,\quad M_\Lambda=e^{K/2}F_\Lambda.
\end{equation}
 In terms of special coordinates the K\"ahler potential is:
\begin{equation}
K(z,\bar z)= - \log (-id_{ABC}(z-\bar z )^A (z-\bar z )^B (z-\bar z )^C).
\end{equation}
The central charge $ Z $ has a form \cite{Cer}:
\begin{equation}
Z=e^{{K(z,\bar z)\over 2}}( X^\Lambda  q_\Lambda  -F_\Lambda
p^\Lambda ).
\end{equation}

 In \cite{BCWKLM} it was proposed to introduce new variables:
\begin{equation}
Y^\Lambda  \equiv \bar Z X^\Lambda .
\end{equation}
Stabilization equations (\ref{stab}) in terms of these new
variables become:
\begin{eqnarray}
ip^\Lambda & = & Y^\Lambda -{\bar Y}^\Lambda \cr
iq_\Lambda & = & F_\Lambda (Y)-{\bar F}_\Lambda (\bar Y).
\label{pqy}\end{eqnarray}
Mass of the double-extreme black hole and entropy then has a form:
\begin{equation}
{S \over \pi }=M_{BPS}^2=|Z|^2=|Y^0|^2 exp (-K(z,\bar z)).
\label{e0}\end{equation}

	In this paper we will show the  solution of equations
(\ref{pqy}) for most
general form of  $d_{ABC}$ and arbitrary $p$ and $q$ charges. In
Section 2 we
will derive the expression for black hole entropy in terms of
$d_{ABC}$ and some
combinations of $ p $ and $ q $. In Section 3 solutions for moduli
fields will
be given and in Section 4 some examples of explicit solutions for entropy for
particular Calabi-Yau
intersection
numbers will be considered.

\section{Entropy of Calabi-Yau black holes.}

   Stabilization equations (\ref{pqy}) and the entropy formula
(\ref{e0}) lead to
the general result for entropy of Calabi-Yau black hole with prepotential
(\ref{ppd}). For the most general case with  $ p^0 \not= 0,
q_0\not= 0 $  and
arbitrary  charges $ (p^A, q_A) $ the entropy of the black hole
depends only on
these charges and  numbers $ d_{ABC} $:
\begin{eqnarray}
{S\over \pi } = {1\over 3p^0}
\sqrt {{4\over 3}{(\Delta _A {\tilde x}^A)}^2 - 9\Bigl(p^0(p\cdot q )- 2
D\Bigr)^2 } \ .
\label{ep1} \end{eqnarray}
Here $ \Delta _A  = 3D_A - p^0q_A  $, and
 $ D_{A}= d_{ABC}p^B p^C $,\
$D  = d_{ABC}p^A p^B p^C$, \ $p\cdot q=p^0q_0+p^Aq_A $.
  In combination
\begin{eqnarray}
{(\Delta _A {\tilde x}^A)}^2 = {\Bigl ((3D_A - p^0q_A) {\tilde
x}^A \Bigr )}^2
\label{dx2} \end{eqnarray}
the variables ${\tilde x}^A$ are the {\it real} solutions of algebraic
system:
\begin{eqnarray}
d_{ABI}{\tilde x}^A {\tilde x}^B =\Delta _I .
\label{alg1} \end{eqnarray}
which is determined by intersection numbers of the Calabi-Yau
manifold and
black hole charges. This system might have no analytical solution
in general
case. The number of equations in (\ref{alg1}) is equal to the
number of moduli
fields $n_v $. Left-hand  parts of each equations are quadratic
forms  with
coefficients $d_{AB(I)}$ and right-hand parts are  arbitrary
and depend
on values of electric and magnetic charges of the black hole. In
other words,
we have  $n_v $  2-dimensional surfaces embedded in  $n_v
$-dimensional space
and intersection of these surfaces is a solution of this system. It
is not
clear yet if there is any connection between the geometry of Calabi-Yau
manifolds and the geometry of this picture. We considered  a few
examples of
Calabi-Yau manifolds with different sets of $d_{ABC}$. For each of
them we
found  the  expression for the black hole entropy. These examples will be
considered in part 4 of this paper.

 Solution of  equations  (\ref{pqy}) becomes much simpler if $
p^0 = 0 $. In
this case we have the system of linear ( instead of quadratic )
equations and there
is a general solution for moduli fields in terms of black hole
charges. The
entropy of the black hole is  then given by the formula:
\begin{eqnarray}
{S\over  \pi}=\sqrt{{D\over 3}(q_B D^B+12q_0)} \ .
\label{ep0} \end{eqnarray}
Here $ D_{AB} = d_{ABC}p^C $,\ $ D^{AB}=[D_{AB}]^{-1}$,\ and $
D^A=D^{AB} q_B$.

It is straightforward to generalize these results to the
prepotential with
additional topological term  $F_1={c_2J_A \over 24} z^A$ .
\begin{equation}
F(X)=d_{ABC}z^Az^Bz^C+\sum_A^h {c_2\cdot J_A\over 24}z^A
\end{equation}
In \cite{BCWKLM} it was shown that the addition of this topological
term leads
to simple transformation of charges $q _{\Sigma }$:
\begin{eqnarray}
{\tilde q}_{ \Sigma }=q_{ \Sigma }-W_{\Sigma \Lambda }p^{\Sigma }\cr
\nonumber\\
W_{0A}={c_2J_A \over 24}.
\end{eqnarray}
Therefore in  our expressions for entropy (\ref{ep1}) and
(\ref{ep0})we should
use these new
${\tilde q}_{ \Sigma } $ connected with old charges by the equations:
\begin{eqnarray}
{\tilde q}_0=q_0-{c_2J_A \over 24}p^{A}\cr
\nonumber\\
{\tilde q}_A=q_A-{c_2J_A \over 24}p^{0}.
\end{eqnarray}

\section{Fixed Moduli solutions}

 Stabilization equations (\ref{pqy}) define values of the
moduli fields
near the black hole horizon through the electric and magnetic
charges of the
black hole. From "$p^{\Lambda }$" equations (\ref{pqy}) it
follows immediately that $
ImY^{\Lambda }={p^{\Lambda }\over 2}$. Therefore the  second part of
(\ref{pqy}),
"$q_{\Lambda }$" -equations, is a system of equations on
$ReY^{\Lambda }$.   In
new variables:
\begin{equation}
x^A=ReY^A-{p^A ReY^0 \over p^0}; \quad x^0=ReY^0
\end{equation}
 this system has a form:
\begin{eqnarray}
 2p^0ReY^0 \Delta _C x^C =
\Bigl(p^0(p\cdot q )- 2 D\Bigl){\bigl( {|Y^0|}^2 \bigr )}^2
\label{x0}\end{eqnarray}
\begin{eqnarray}
d_{ABI}x^A x^B ={\Delta _I \over 3{p^0}^2}{|Y^0|}^2 .
\label{xi}\end{eqnarray}
where $
|Y^0|^2={(ReY^0)}^2+{{p^0}^2\over 4}$.
Solution of equations (\ref{x0})and (\ref{xi}) can be expressed in
terms of
variables ${\tilde x}^A $ from (\ref{alg1}), in that case :
\begin{eqnarray}
x^A={\tilde x}^A \sqrt{{ {|Y^0|}^2 \over 3{p^0}^2}}
\end{eqnarray}
and
\begin{eqnarray}
|Y^0|^2 = {{p^0}^2\over 3}{ {(\Delta _C {\tilde x}^C)}^2 \over
{4\over 3}{(\Delta _C {\tilde x}^C)}^2 - 9(p^0(p\cdot q )- 2 D)^2 }.
\end{eqnarray}
Finally the  moduli fields $z^A$ are:
\begin{eqnarray}
z^A = {3\over 2}{ {\tilde x}^A \over p^0 (\Delta _C {\tilde
x}^C)}(p^0(p\cdot q
)- 2 D)+
{p^A\over p^0}- i{3\over 2}{ {\tilde x}^A \over  (\Delta _C {\tilde
x}^C)}{S\over \pi }
\end{eqnarray}
 where the  entropy $ S/\pi $ is given by (\ref{ep1}) and
${\tilde
x}^A$ are  solutions of algebraic system:
\begin{eqnarray}
d_{ABI}{\tilde x}^A {\tilde x}^B =\Delta _I .
\label{alg} \end{eqnarray}

   In case $p=0$ equations the system (\ref{pqy})  becomes  very simple:
\begin{eqnarray}
6D_{AI}ReY^A =q_IReY^0 \cr
\nonumber\\
3D_{AB} ReY^A ReY^B-{D\over 4}=-q_0(ReY^0 )^2
\end{eqnarray}
and the solution for moduli is:
\begin{eqnarray}
z^A={D^A\over 6}-i{p^A\over 2}\sqrt{{q_B D^B+12q_0\over 3D}}={D^A\over
6}-i{p^A\over 2}{DS\over \pi }
\end{eqnarray}
were $S/\pi $ is given by  (\ref{ep0}).

\section{Examples.}

We will present below some examples of the entropy formulas for
particular CY
manifolds.

\quad  In case  of  two moduli $ n_v=2 $ the   algebraic system
(\ref{alg1}) has
a general solution.  Expression $ (\Delta _C {\tilde x}^C)^2 $ in entropy
formula (\ref{ep1}) has the form:
\begin{eqnarray}
(\Delta _C {\tilde x}^C)^2&=&\Bigl ( -2\Delta _A \Delta _B \Delta _C
d_{A}^{BC}\det
d_1 \det
d_2\cr
\nonumber\\
&+&(\det \hat d_{123})\bigl (\Delta _1 \Delta _A \Delta _B d_{2}^{AB}\det
d_2 +\Delta
_2 \Delta _A \Delta _B d_{1}^{AB}\det d_1 \bigr ) \cr
\nonumber\\
&+&2\bigl ( -{\Delta _1}^2\det d_2 - {\Delta _2}^2\det d_1+\Delta _1\Delta
_2(\det \hat d_{123})\bigr )^{3/2} \Bigr )\cdot \Bigl ( (\det \hat
d_{123})-4\Delta _1\Delta
_2 \Bigr
)^{-1}
\end{eqnarray}
were $ (\det \hat d_{123})=d_{111}d_{222}- d_{112}d_{122}$
and
\begin{eqnarray}
\det d_1 = \det( d_{1AB})\not=  0, \  \det d_2 = \det (d_{2AB})\not=  0,
\label{d=0} \\
d_1^{AB} = [d_{1AB}]^{-1},\quad  d_2^{AB} = [d_{2AB}]^{-1}.
\end{eqnarray}
Even if these conditions are not satisfied and there is no inverse
matrix $
d_1^{AB} $ or  $ d_1^{AB} $ a solution can still exist.
One of particular choices of Calabi-Yau intersection numbers with
$ n_v = 2 $
for which condition (\ref{d=0}) does not work was given in
\cite{HLY}. In that
case:
\begin{equation}
d_{111}=8/6;\quad d_{112}=4/6;\quad d_{122}=d_{222}= 0.
\end{equation}
Solution of  (\ref{alg1}) exists and expression (\ref{dx2}) has a form:
\begin{eqnarray}
(\Delta _C {\tilde x}^C)^2=3/8 \Delta _2 \bigl ( 3 \Delta _1
-2\Delta _2\bigr
)^2
\end{eqnarray}
so that  black hole entropy is:
\begin{eqnarray}
{S\over \pi } = {1\over 3p^0}
\sqrt{{\Delta _2 \over 2}( 3 \Delta _1 -2\Delta _2 )^2 -
9(p^0(p\cdot q )- 2
D)^2 }.
\end{eqnarray}

   One of examples of model with $ n_v = 3 $ is  STU model studied in
\cite{WCLMR} with prepotential  $ F(X)=d_{ABC}z^Az^Bz^C $ and $
d_{ABC}=d_{123}=1/6 $ .
 Expressions for entropy and for moduli fields were already found in
\cite{BKRSW}.  Here we show these expressions in terms of solutions of
algebraic system (\ref{alg1}) and $ \Delta _A $:
\begin{eqnarray}
(\Delta _C {\tilde x}^C)^2={9\over 2}{\Delta _1 \Delta _2 \Delta _3 \over
d_{123}}.
\end{eqnarray}
and entropy  (\ref{ep1}) is
\begin{eqnarray}
{S\over \pi } = {1\over p^0}
\sqrt {{2\over 3d_{123}}\Delta _1 \Delta _2 \Delta _3  -
(p^0(p\cdot q )- 2
D)^2 }
\end{eqnarray}
this expression coincides with expression for entropy of STU black
holes in
\cite{BKRSW}.
Expression for moduli is:
\begin{eqnarray}
z^A = { 3(p^0(p\cdot q )- 2 D)+6p^A \Delta _A\over 2p^0 \Delta _A}-
i{S\over
2\pi \Delta _A}
\quad (no\ summation\ on\ A)
\end{eqnarray}
which also  coincides with expression for moduli from \cite{BKRSW}.

     Some interesting examples of  $ n_v = 3 $  Calabi-Yau
manifolds were given
in  \cite{LSTY}.
Those  examples of classical prepotentials correspond
to Type II
vacuum. Each of  prepotentials are related with each other and
with  $
F_{II}^0= STU + {1\over 3 } U^3 $ prepotential by linear transformations:
\begin{eqnarray}
F_{II}^1 = {4\over 3}{(z^1)}^3+{(z^1)}^2z^2+{(z^1)}^2z^3 + z^1z^2z^3 \\
(z^1=U,
z^2=T-U, z^3=S-U);
\label{F1}\end{eqnarray}
\begin{eqnarray}
F_{II}^2 = {4\over 3}{(z^1)}^3+2{(z^1)}^2z^2+ z^1{(z^2)}^2+{(z^1)}^2z^3 +
z^1z^2z^3 \ ,\\
(z^1=U, z^2=T-U,
z^3=S-T)\ ;
\label{F2}\end{eqnarray}
\begin{eqnarray}
F_{II}^3 = {4\over 3}{(z^1)}^3+{2\over 3}{(z^1)}^2z^2+ {1\over
2}z^1{(z^2)}^2+{(z^1)}^2z^3 + z^1z^2z^3\ , \\
(z^1=U, z^2=T-U,
z^3=S-1/2T-1/2U) \ .
\label{F3}\end{eqnarray}
   Solution of algebraic equation (\ref{alg1}) exists for each of
these prepotentials and the entropy for  $ F_{II}^0 $ is:
\begin{eqnarray}
{S\over \pi } & = &{1\over 3p^0}
\sqrt {{4\over 3}\Bigl ({3\over 2}\Delta _3 \bigl ((\Delta _3
)^2+12\Delta _1 \Delta _2 \bigr ) + {3\over 2} (W_0)^{{3\over
2}}\Bigr ) -
9\Bigl( p^0(p\cdot q )- 2
D\Bigr)^2 } \label{s0} \\
\nonumber\\
W_0& = &\Bigl( (\Delta _3 )^2-4\Delta _1 \Delta _2 \Bigr),
 \end{eqnarray}
expressions  $ (\Delta _C {\tilde x}^C)^2 $ for $ F_{II}^1$, $
F_{II}^2 $ and $
F_{II}^3$ are:
\begin{eqnarray}
(\Delta _C {\tilde x}^C)^2_{F1}& = &{3\over 2}(\Delta _1 -\Delta _2
-\Delta _3
)\bigl ((\Delta _1 -\Delta _2 -\Delta _3 )^2+12\Delta _2 \Delta _3
\bigr ) +
{3\over 2} (W_1)^{{3\over 2}}, \cr
\nonumber\\
W_1& = &\Bigl( (\Delta _1 -\Delta _2 -\Delta _3 )^2-4\Delta _2 \Delta _3 \Bigr)
{}.
\label{s1}\end{eqnarray}
\begin{eqnarray}
(\Delta _C {\tilde x}^C)^2_{F2}& = &{3\over 2}(\Delta _1 -\Delta _2
)\bigl
((\Delta _1 -\Delta _2 )^2+12\Delta _3(\Delta _2- \Delta _3) \bigr
) + {3\over
2} (W_2)^{{3\over 2}}, \cr
\nonumber\\
W_2& = &\Bigl( (\Delta _1 -\Delta _2)^2-4\Delta _3(\Delta _2- \Delta _3)
\Bigr).
\label{s2}\end{eqnarray}
\begin{eqnarray}
(\Delta _C {\tilde x}^C)^2_{F3}& = &{3\over 2}(\Delta _1 -\Delta _2
- {1\over
2}\Delta _3 )\bigl ((\Delta _1 -\Delta _2 -{1\over 2}\Delta _3
)^2+12\Delta _3
(\Delta _2- {1\over 2}\Delta _3)\bigr )\cr
\nonumber\\
 & + &{3\over 2} (W_3)^{{3\over
2}},  \cr
\nonumber\\
W_3& = &\Bigl( (\Delta _1 -\Delta _2 -{1\over 2}\Delta _3 )^2-4\Delta _3
(\Delta _2-
{1\over 2}\Delta _3) \Bigr) .
\label{s3}\end{eqnarray}
Solutions  (\ref{s1}), (\ref{s2}) and (\ref{s3}) are connected with
each other
and with  $ (\Delta _C {\tilde x}^C)^2 $ solution for $ F_{II}^0$ by
linear transformations of $\Delta _A $. For example, connection between
(\ref{s1}) and (\ref{s2}) is :  $ \{ \Delta _1,\Delta _2,
\Delta _3 \}
\rightarrow \{ \Delta _1,(\Delta _2-\Delta _3), \Delta _3 \}$ and between
(\ref{s1}) and (\ref{s3}) is :
$ \{ \Delta _1,\Delta _2, \Delta _3 \} \rightarrow \{ \Delta _1,(\Delta
_2-{1\over 2}\Delta _3), \Delta _3 \} $.

The last example of  Calabi-Yau manifolds that we are going to
consider here
was again given in \cite{LSTY}. This model is connected with $ F=
{3\over 8} U^3
+{1\over 2} UT^2-{1\over 6} V_X^3$ by linear transformation:
\begin{eqnarray}
z^1=V_X, \quad z^2=U-V_X,\quad  z^3= T-{3\over 2 }U.
\end{eqnarray}
Classical prepotential in this case is:
\begin{eqnarray}
F & = & {4\over 3}{(z^1)}^3+{3\over 2}{(z^2)}^3+ {9\over 2}{(z^1)}^2z^2+{9\over
2}z^1{(z^2)}^2
+{3\over 2}{(z^1)}^2z^3 \cr
\nonumber\\
& + &{3\over 2}{(z^2)}^2z^3 + {1\over 2}z^1{(z^3)}^2 + {1\over 2}z^2{(z^3)}^2 +
3z^1z^2z^3.
\end{eqnarray}
the solution for (\ref{ep1}) is:
\begin{eqnarray}
{S\over \pi } = {1\over 3p^0}\,
\sqrt {{4\over 3}{\left( \sqrt {6(\Delta _2 -\Delta
_1)^3}+ \sqrt
{{2\over 3}(\Delta _2 -3\Delta _3)^3}-\sqrt {{2\over 3}(\Delta
_2)^3}\  \right)}^2 - 9\Bigl(p^0(p\cdot q )- 2
D\Bigr)^2 }
 \end{eqnarray}
We have suppressed here the stabilized values of moduli fields although they
are
available for each case.  The expressions for entropy of black holes  are
rather complicated, but they could be compared with
entropy obtained by counting of microscopic  degrees of freedom ( stringy
microstates, "D-branes").
\vskip 1 cm

 In this paper we have found the entropy-area formula of N=2
extreme black hole
 for most
general case of Calabi-Yau moduli space. We have shown that the
entropy and the
moduli fields in general case
 depend on the solution of algebraic system of quadratic equations
(\ref{alg1}).
We have found the entropy of the black hole for some particular
examples of
Calabi-Yau moduli space.
Although it is not obvious  that the existence of the solutions can  be
regarded as  general property of  Calabi-Yau
prepotentials we hope  that there is a connection between  them.
It will be
very interesting to find such a connection.

\section{Acknowledgments.}
  Author is grateful to Renata Kallosh for stimulating discussions
and help.
Useful discussions with Soo-Jong Rey, Wing Kai Wong and Arvind
Rajaraman are
gratefully
acknowledged. The work  is supported by the Department of Energy
under contract
DOE-DE-FG05-91ER40627 and the NSF grant  PHY-9219345.

\end{document}